\documentclass[aps,onecolumn,epsfig,prb]{revtex4}
\usepackage{graphicx}

\begin{document}

\title{Phase space topology of a switching current detector}

\author{Joachim Sj\"ostrand, Hans Hansson, Anders Karlhede}
\affiliation{Department of Physics, Stockholm University, AlbaNova University Center, SE-10691 Stockholm, Sweden}

\author{Jochen Walter, Erik Thol$\acute{e}$n, David Haviland}
\affiliation{Nanostructure Physics, Royal Institute of Technology,AlbaNova University Center, SE-10691 Stockholm, Sweden}

\date{\today}

\begin{abstract}
\noindent
We examine in theory and by numerical simulation, the dynamic process of switching from a zero voltage to a finite voltage state in a Josephson junction circuit. The theoretical model describes small capacitance Josephson junctions which are overdamped at high frequencies, and can be applied to detection of the quantum state of a qubit circuit. We show that the speed and fidelity of the readout are strongly influenced by the topology of the phase space attractors. The readout will be close to optimal when choosing the circuit parameters so as to avoid having an unstable limiting cycle which separates the two basins of attraction.
\end{abstract}

\maketitle

\vskip0.1pc


One of the unique and exciting characteristics of quantum bits based on Josephson junction circuits (as opposed to quantum systems like atoms or nuclear spins) is the ability to engineer the system for optimal performance with regard to decoherence and readout efficiency. With Josephson junction circuits we have electrical access to the Hamiltonian of a quantum system described by two conjugate circuit variables.  External current sources or voltage sources can be used to change potential and kinetic energies in the Hamiltonian of the quantum circuit\cite{MSS}. This fact allows one to design a fast and reliable readout.
 
The quantum dynamics of Josephson junction circuits is intensively studied by several groups around the world today. Many different circuits and measurement schemes have been proposed and implemented in experiments which prepare, evolve, and read out the quantum state of the circuit\cite{Nakamura,Vion,Chirescu,Martinis,Duty}. One of the most successful readout methods used this far is based on the switching of a Josephson junction from the zero voltage state to the finite voltage state. In this letter we analyze this switching process in some detail, focusing on the reliability of the detection method in the presence of external noise. We show that the existence of an unstable limiting cycle in the phase space dynamics of the Josephson junction leads to late retrapping (long measurement time) and false switching events, and that such an unstable cycle can be avoided with proper choice of parameters, while maintaining the desired overdamped phase dynamics.  

The switching current detector works on the principle that the quantum states one wants to differentiate have different critical currents $I_{0n}=max[dE_n/d\phi]$, where $E_n$ is the energy of eigenstate $n$, and $\phi$ is the external phase of the circuit, which can be changed by application of an external current. Restricting ourselves to the two lowest energy states of the circuit (a qubit) we quickly ramp the current to a value $I_p$, where $I_{00}>I_p>I_{01}$. The circuit will evolve differently, depending on whether it is in state 0 or in state 1. If the qubit is in state 0, the circuit will not switch, meaning that it remains in the zero voltage state where the phase is trapped in a local minimum of the potential $E_0(\phi)$. If the qubit is in state 1, it will switch, meaning that the phase escapes from this minimum, and evolves to a "free running" state with finite voltage.  Reading the absence or presence of this voltage constitutes a determination of the quantum state.

A good switching detector should have a bistability, characterized by two stable attractors in the phase space of the non-linear circuit. As we shall explain, such a bistability makes possible a "sample and hold"  approach to quantum measurement\cite{JW}. For the Josephson junction qubit detectors studied thus far, bistability has been realized by switching Josephson junctions with underdamped phase dynamics at all frequencies. A junction with overdamped dynamics at all frequencies would not have a bistability. Overdamped dynamics is however desirable from the point of view of minimal phase fluctuations (which cause dephasing of the qubit) and rapid reset time of the switching detector. Therefore, an interesting model in the context of qubit detectors is the Josephson junction shunted by a series RC circuit, which has the property of overdamped phase dynamics at high frequencies, and underdamped dynamics at low frequencies. This model has been studied by several authors\cite{Ono87,KM90,Vion2,Joyez,JS}. Kautz and Martinis showed that bistability is possible in this model, with one state characterized by ``phase diffusion'' where the phase is thermally kicked between minima of the potential, and the other state is that of a free running phase.

A good detector has high readout fidelity and short measurement time. To achieve high fidelity the circuit should be insensitive to all sources of noise which may cause false switching events, so that the switching probability is only determined by the quantum probabilities. Short measurement time is necessary so that the measurement occurs before the qubit has time to change state (relaxation or excitation). In the switching detectors studied thus far, the fidelity and speed is thought to be limited by macroscopic quantum tunneling of the phase in the Josephson potential \cite{Ith,Ank}. Here we consider switching of junctions which are very overdamped at high frequencies, with small capacitance and shunting impedance much less than the quantum resistance ($h/e^2$). In this case a classical description of the non-linear phase dynamics is adequate.

\begin{figure}[tbp]
\begin{center}
\includegraphics[angle=0,width=0.43\textwidth]{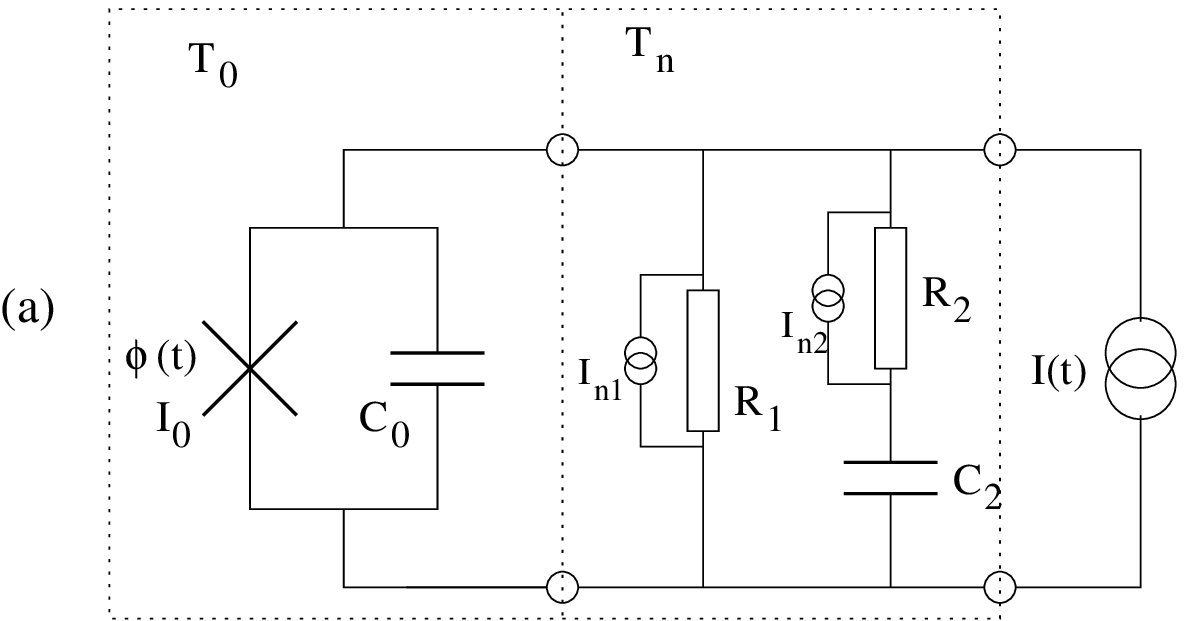}
\includegraphics[angle=0,width=0.43\textwidth]{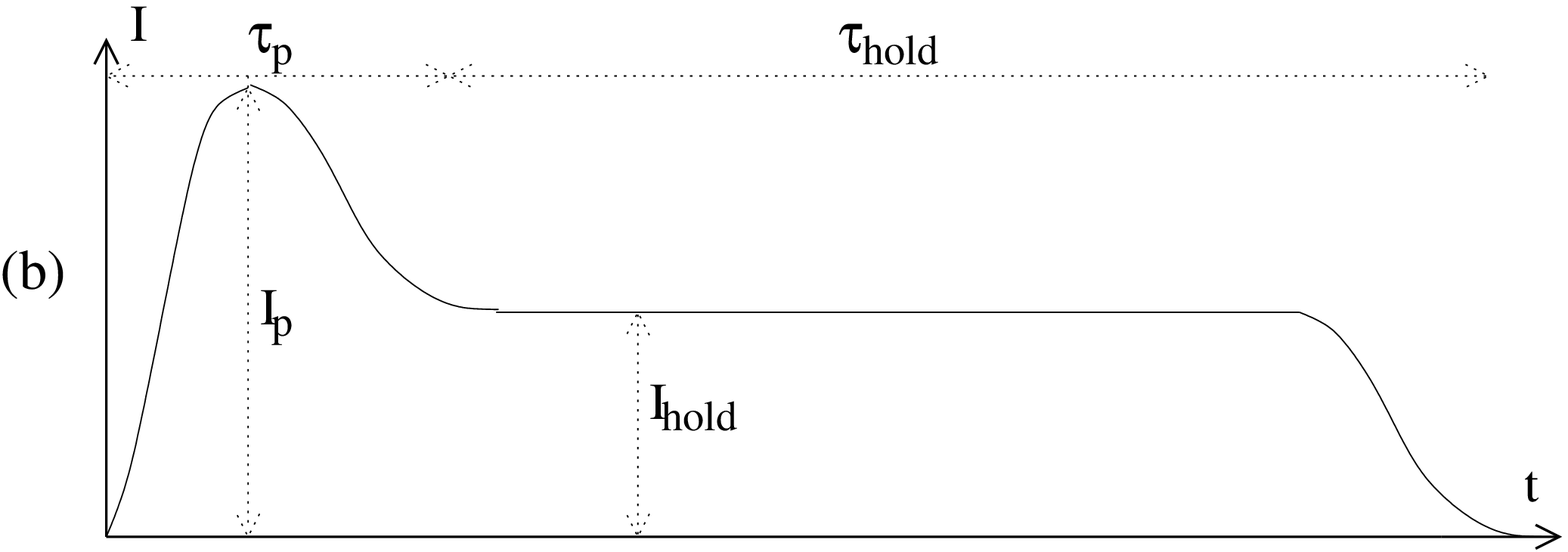}
\caption{The model circuit (a) and the applied current pulse (b).}
\label{Setup}
\end{center}
\end{figure}

The most simple qubit circuit is a Cooper pair transistor. However, in order to study the switching process it is sufficient to consider only a single small capacitance Josephson junction.  The Josephson junction has critical current $I_0$, capacitance $C_0$ and is biased by a current source $I(t)$.  In parallel with the junction is a resistor $R_1$ and a series combination of a resistor $R_2$ and a capacitor $C_2$. See Fig.~\ref{Setup}(a). 

The classical dynamics of our system can be described in terms of $\phi$ (the phase over the junction), $v$ and $v_C$ (the ratio of the voltages over the junction and over $C_2$ to $R_1 I_0$). However, for the small junctions considered here, the current through $C_0$ can to a very good approximation be neglected and we are left with
\begin{eqnarray}
\label{Deq_Zrrc}
\dot \phi & = & \frac{Q_1}{Q_0} \left[ i - \sin \phi + v_C (Q_0/Q_1-1) + i_{n1} +i_{n2} \right] \nonumber \\
\dot v_C & = & \frac{\rho Q_1}{Q_0^3} \left[ i  - \sin \phi - v_C + i_{n1} +i_{n2}\frac{Q_1^2}{Q_0(Q_0-Q_1)} \right].
\end{eqnarray}
Here $i=I/I_0$, $i_{n1,2}$ are noise currents (which are Gaussian distributed and obey $\langle i_{n1,2}(t_1') i_{n1,2}(t_2') \rangle = 2 \frac{k_B T_{n1,2}}{E_J} R_1/R_{1,2} \delta(t_2' - t_1')$ where $t'=2eR_1 I_0t/\hbar$, $Q_0= R_1 \sqrt{2eI_0 C_0 / \hbar}$ is the low frequency quality factor, $Q_1= (R_1^{-1}+R_2^{-1})^{-1}\sqrt{2eI_0 C_0 / \hbar}$ is the high frequency quality factor and $\rho=R_1 C_0 /R_2 C_2$. The combination $(i-\sin \phi)$ is the force associated with the well known tilted washboard potential.

\begin{figure}[htbp]
\begin{center}
\includegraphics[angle=0,width=0.45\textwidth]{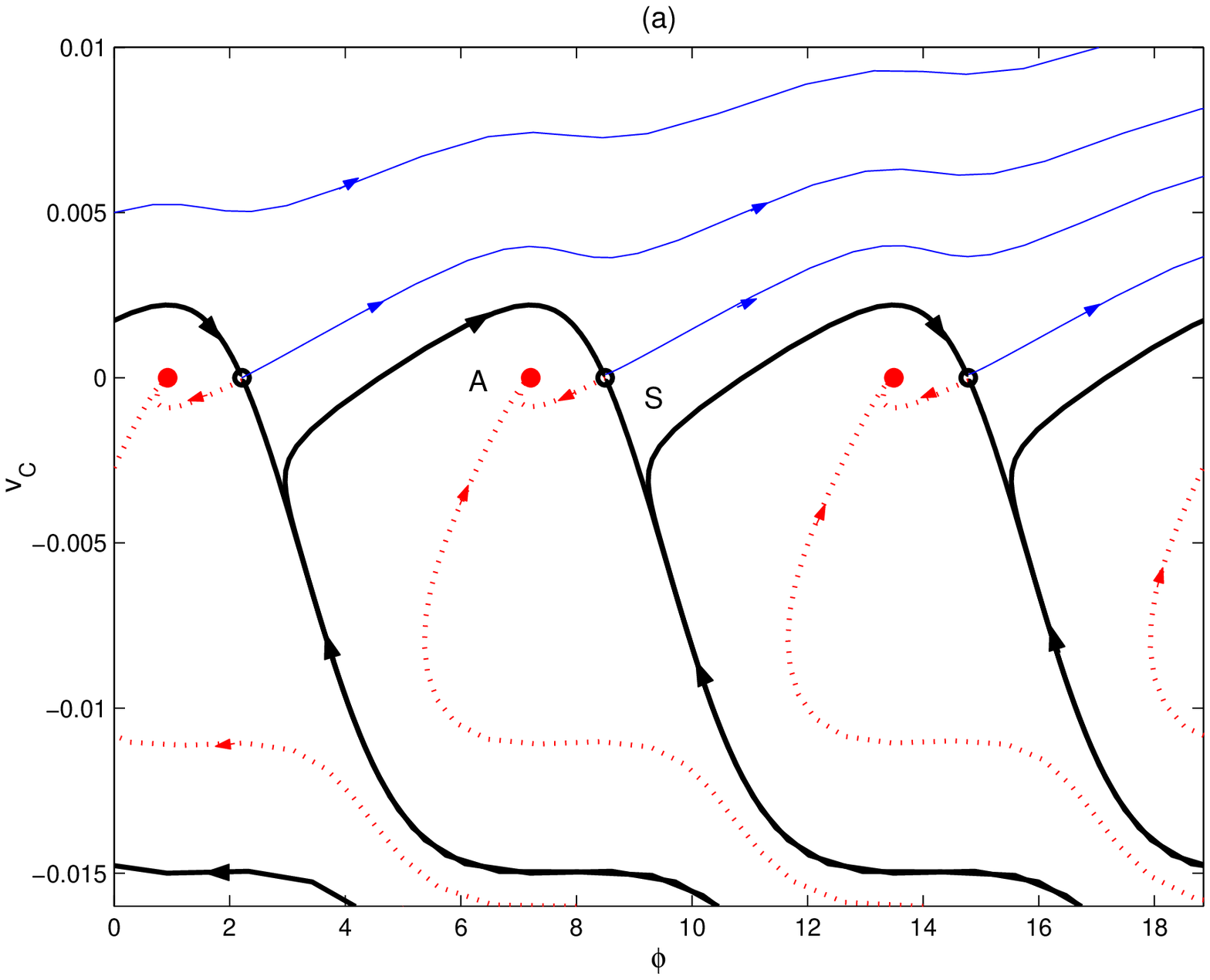}
\includegraphics[angle=0,width=0.45\textwidth]{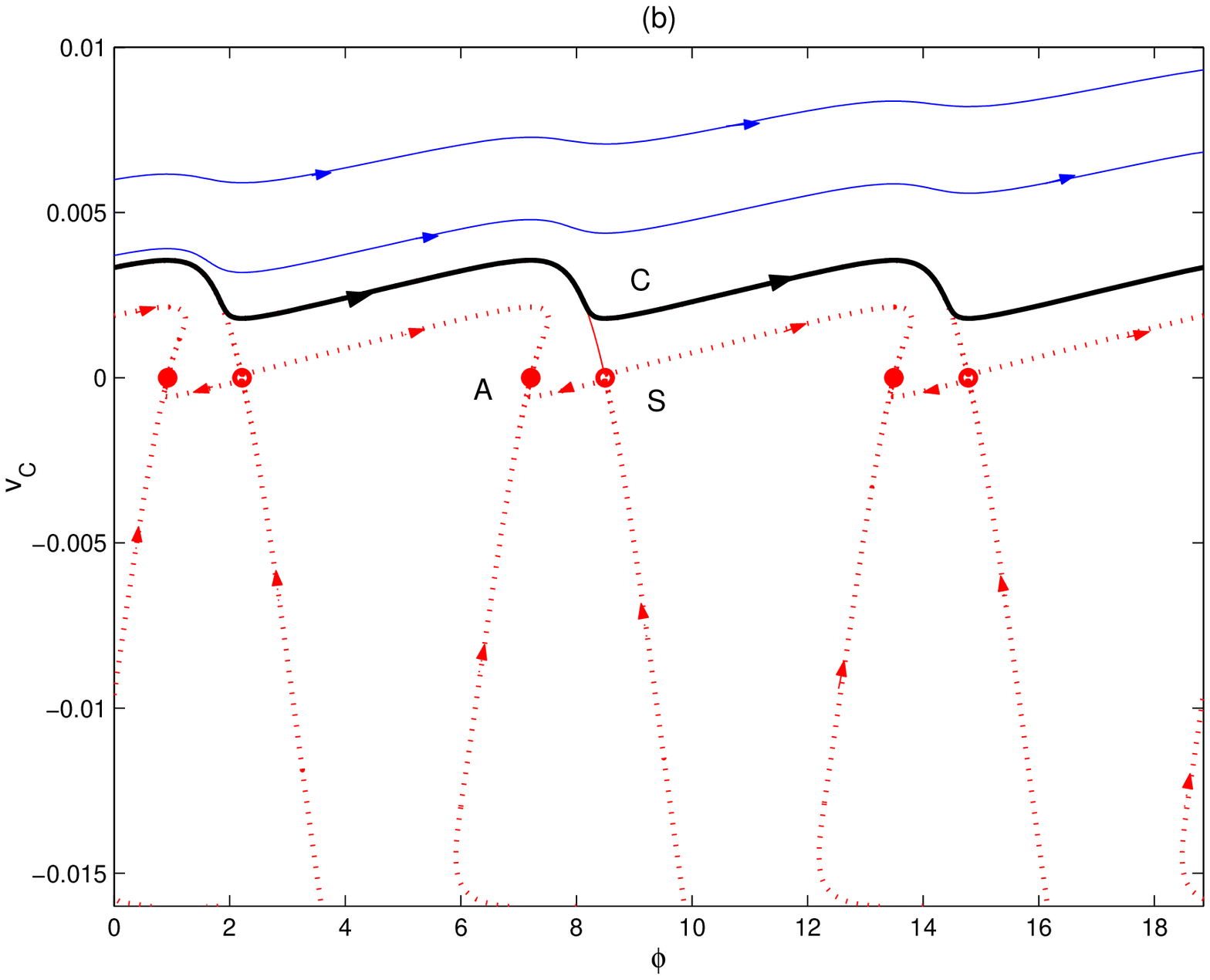}
\caption{(Color online) ($\phi$,$v_C$) phase space flows for $Q_0=3.5$, $i_{hold}=0.80$ and (a) $Q_1=0.02$, $\rho=0.014$; (b) $Q_1=0.04$, $\rho=0.006$.}
\label{Phase_flows}
\end{center}
\end{figure}

Fig.~\ref{Phase_flows} shows the $(\phi,v_C)$ phase portrait for a finite bias current $i_{hold}=0.80$ which is applied during readout. Two cases are shown, both having $Q_0=3.5$, with (a) $Q_1=0.02$ and $\rho=0.014$ and (b) $Q_1=0.04$ and $\rho=0.006$. In both cases there are stable fix points A (located at $(\arcsin(i_{hold})+2\pi n,0)$ corresponding to local minima of the washboard potential; shown with filled red circles), and saddle points S (located at $(\pi-\arcsin(i_{hold})+2\pi n,0)$ corresponding to local maxima of the washboard potential; shown with red/black cirlces). There is also a stable limiting cycle B (located close to $v_C \sim i_{hold}$, not shown in the figures). The regions of dotted red trajectories are the basins of attraction for the different A's (non-switching events), the region of solid blue trajectories is the basin of attraction for B (switching events). The boundary between the switching and the non-switching regions are shown with thick black trajectories. The arrows point in the direction of positive time evolution. Figs.~\ref{Phase_flows}(a) and (b) show a striking difference in topology: in (b) the boundary is an unstable limiting cycle C (with $0<v_C<v_{sw}$ being periodic in $\phi$), whereas in (a) there are several boundaries (all ending in the different S's) which separate the different red regions. As $Q_1$ is decreased, C approaches S. When C collides with S and vanishes, the black trajectory flips origin to $v_C=-\infty$, and the blue basin stretches down and separates the different red basins as shown in (a). For the parameters used here, this flip occurs close to $Q_{1c}\approx0.029$.

The sample and hold switching measurement is realized by a fast initial pulse $I_p$, with duration $\tau_p$ as shown in Fig. 1(b). After this pulse, the phase space coordinate $(\phi(\tau_p),v_C(\tau_p))$ will lie close to the boundary between the basins of attraction. The existence of the unstable boundary C imples two things: first, a long time is required for trajectories near the boundary to separate into the two basins of attraction. Second, events can jump between consecutive (red) non-switching basins without passing the blue switching region. When $Q_1$ increases, both the separation time and the distance between A and C will increase. Hence for $Q_1>Q_{1c}$ (Fig.~\ref{Phase_flows}(b)), the topology gives rise to ``late retrappings'' and ``late switches'' which increases the measurement time and result in errors. On the other hand, for $Q_1<Q_{1c}$ (Fig.~\ref{Phase_flows}(a)) the boundary is a saddle point flow, and there is no C, and consequently no late retrapping or late switching. At finite temperatures late events are still possible, although the probability of occurrence is much smaller than for $Q_1>Q_{1c}$ and decreases with smaller $Q_1$. 

\begin{figure}
\begin{center}
\includegraphics[angle=0,width=0.45\textwidth]{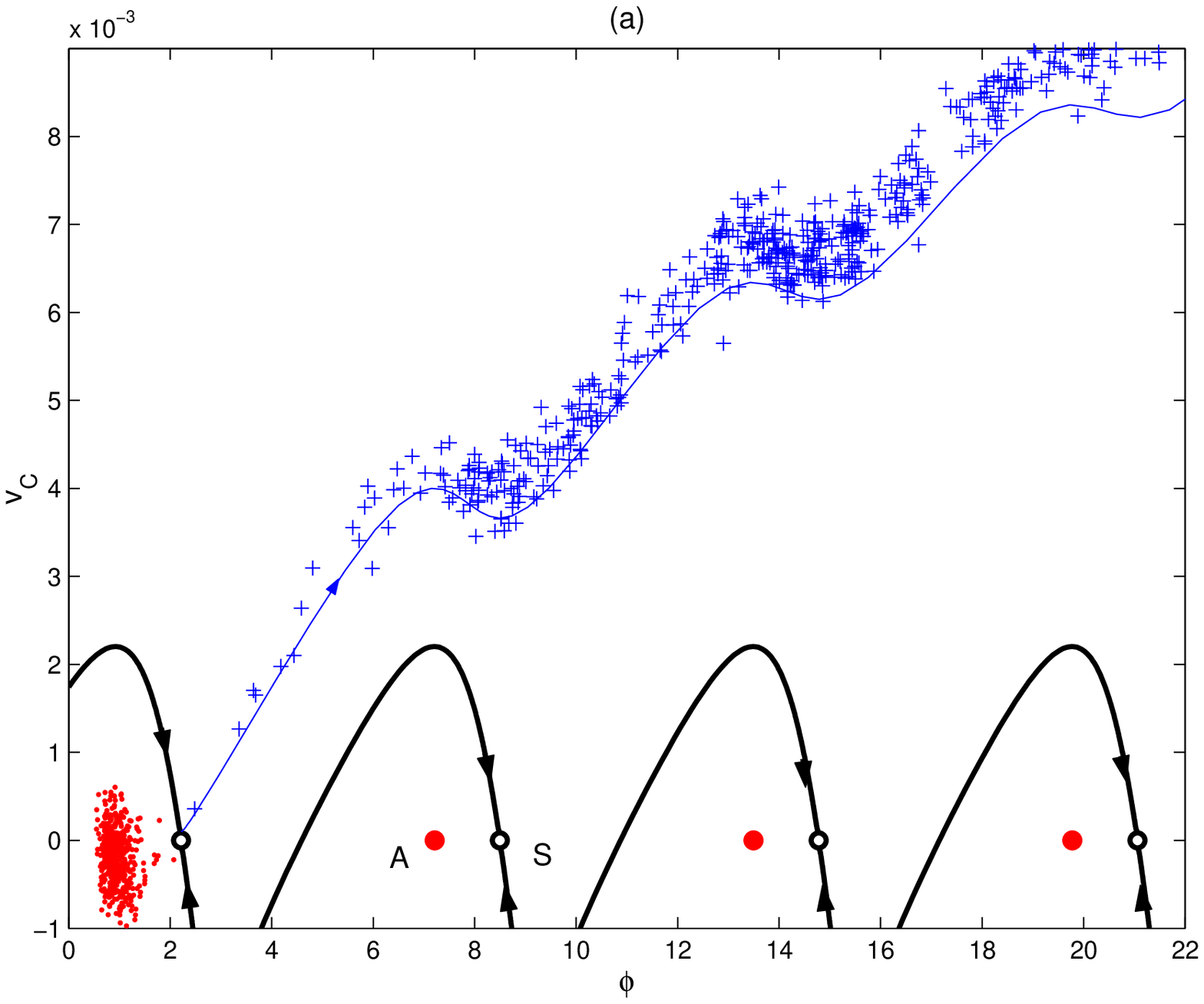}
\includegraphics[angle=0,width=0.45\textwidth]{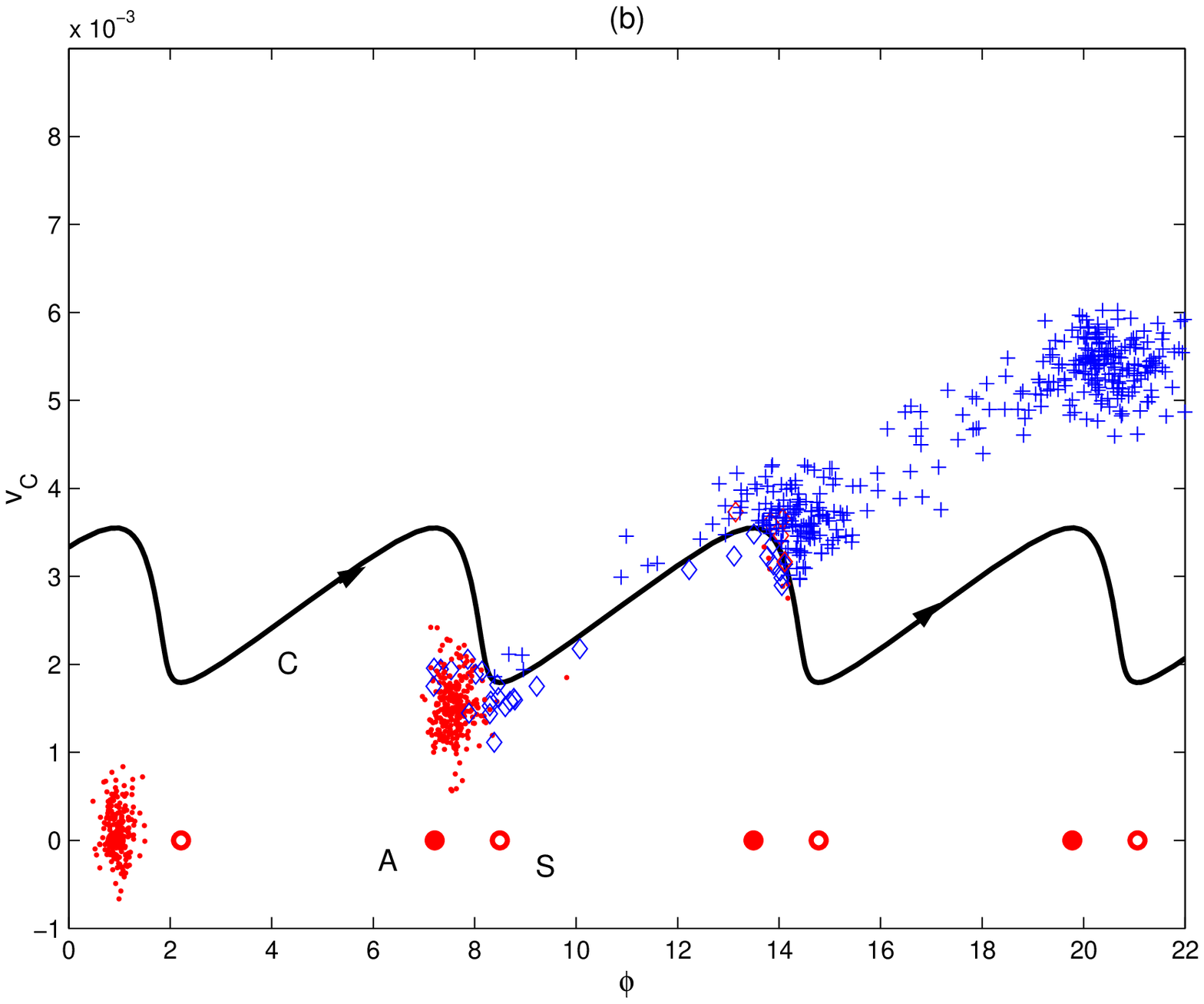}
\caption{(Color online) Simulated $(\phi,v_C)$ values at $t=6.5$ ns for 1000 events, $i_p$ giving 50\% switching, $Q_0=3.5$, $i_{hold}=0.80$ and $T_n=30$ mK. Switching (non-switching) events are marked with blue plus-signs (red dots), miscounts with diamonds. In (a) $Q_1=0.02$, $\rho=0.014$ and in (b) $Q_1=0.04$, $\rho=0.006$.}
\label{Num_Sim_ufi}
\end{center}
\end{figure}

To demonstrate this sensitivity to phase space topology we have carried out numerical simulations with experimentally relevant parameters; $\tau_p=5$ ns, $I_0=100$ nA, $C_0=10$ fF, $T_n=30$ mK, and a 2\% random variation in the value of $I_p$ which gives 50\% switching. Initially, $(\phi(0),v_C(0))=(0,0)$ which corresponds to the phase being trapped in a minimum of the washboard. Adding the current pulse and hold and thermal noise will evolve the system to a distribution $\{(\phi(t),v_C(t))\}$. The speed of the detector is determined by the time $\tau_m$ at which the distribution distinctly separates into the two basins of attraction. In the optimal case, $\tau_m=\tau_p$.

In Fig.~\ref{Num_Sim_ufi}(a,b) we have plotted ($\phi$,$v_C$) distributions for $1000$ events at $t=6.5$ ns. Switching (non-switching) events are marked with blue plus-signs (red dots), events that change basin after $6.5$ ns (i.e. miscounts) are marked with diamonds in the respective color. We see that the distribution for $Q_1=0.02$ (topology of Fig.~\ref{Phase_flows}(a)) is clearly separated into the two basins of attraction, with only a handfull of events close to the boundary, and no miscounts. Note that all non-switching events are in the first minimum, and all switching events emanate from the first saddle point, with no late switches or late retrappings. On the other hand, for $Q_1=0.04$, many events lie close to C and there are several late switches and late retrappings. There are 34 miscounts, and for higher $Q_1$ there will be many more such events.

In summary, we have shown that the phase space topology plays a crucial rule in the fidelity and speed of switching current readout for quantum bits. We find that it is possible to achieve almost optimal speed and very high fidelity with overdamped phase dynamics at high frequencies.

We would like to acknowledge helpful conversations with D. Vion and S. \"Ostlund. This work was partially supported by the EU project SQUBIT-2.


\end{document}